\documentclass[12pt]{article}
\usepackage{amsmath,amsfonts,amssymb}

\begin{document}
\thispagestyle{empty}

\def\theequation{\arabic{section}.\arabic{equation}}
\def\a{\alpha}
\def\b{\beta}
\def\g{\gamma}
\def\d{\delta}
\def\dd{\rm d}
\def\e{\epsilon}
\def\ve{\varepsilon}
\def\z{\zeta}
\def\B{\mbox{\bf B}}

\newcommand{\h}{\hspace{0.5cm}}

\begin{titlepage}
\vspace*{1.cm}
\renewcommand{\thefootnote}{\fnsymbol{footnote}}
\begin{center}
{\Large \bf More three-point correlators of} \vskip 0cm {\Large\bf
giant magnons with finite size}
\end{center}
\vskip 1.2cm
\centerline{\bf Plamen Bozhilov}
\vskip 0.6cm
\centerline{\sl Institute for Nuclear Research and Nuclear Energy}
\centerline{\sl Bulgarian Academy of Sciences} \centerline{\sl  1784
Sofia, Bulgaria}

\centerline{\tt plbozhilov@gmail.com}

\vskip 20mm

\baselineskip 18pt

\begin{center}
{\bf Abstract}
\end{center}
\h In the framework of the semiclassical approach, we compute the
normalized structure constants in three-point correlation functions,
when two of the vertex operators correspond to heavy string states,
while the third vertex corresponds to a light state. This is done
for the case when the heavy string states are {\it finite-size}
giant magnons with one or two angular momenta, and for two different
choices of the light state, corresponding to dilaton operator and
primary scalar operator. The relevant operators in the dual gauge
theory are $Tr\left(F_{\mu\nu}^2 \ Z^j+\ldots\right)$ and $Tr\left(
Z^j\right)$. We first consider the case of $AdS_5\times S^5$ and
$\mathcal{N} = 4$ super Yang-Mills. Then we extend the obtained
results to the $\gamma$-deformed $AdS_5\times S^5_\gamma$, dual to
$\mathcal{N} = 1$ super Yang-Mills theory, arising as an exactly
marginal deformation of $\mathcal{N} = 4$ super Yang-Mills.

\end{titlepage}
\newpage

\def\nn{\nonumber}
\def\tr{{\rm tr}\,}
\def\p{\partial}
\newcommand{\bea}{\begin{eqnarray}}
\newcommand{\eea}{\end{eqnarray}}
\newcommand{\bde}{{\bf e}}
\renewcommand{\thefootnote}{\fnsymbol{footnote}}
\newcommand{\be}{\begin{equation}}
\newcommand{\ee}{\end{equation}}

\vskip 0cm

\renewcommand{\thefootnote}{\arabic{footnote}}
\setcounter{footnote}{0}


\setcounter{equation}{0}
\section{Introduction}

It is known that the correlation functions of any conformal field
theory can be determined  in principle in terms of the basic
conformal data $\{\Delta_i,C_{ijk}\}$, where $\Delta_i$ are the
conformal dimensions defined by the two-point correlation functions
\begin{equation}\nn
\left\langle{\cal O}^{\dagger}_i(x_1){\cal O}_j(x_2)\right\rangle=
\frac{C_{12}\delta_{ij}}{|x_1-x_2|^{2\Delta_i}}
\end{equation}
and $C_{ijk}$ are the structure constants in the operator product
expansion
\begin{equation}\nn
\left\langle{\cal O}_i(x_1){\cal O}_j(x_2){\cal
O}_k(x_3)\right\rangle=
\frac{C_{ijk}}{|x_1-x_2|^{\Delta_1+\Delta_2-\Delta_3}
|x_1-x_3|^{\Delta_1+\Delta_3-\Delta_2}|x_2-x_3|^{\Delta_2+\Delta_3-\Delta_1}}.
\end{equation}
Therefore, the determination of the initial conformal data for a
given conformal field theory is the most important step in the
conformal bootstrap approach. While this is well-established in two
dimensions where the conformal symmetry is infinite dimensional
\cite{BPZ}, it is extremely difficult to extend the procedure to
higher dimensional  space-times.

The AdS/CFT correspondence \cite{AdS/CFT} between type IIB string
theory on ${\rm AdS}_5\times S^5$ space and ${\cal N}=4$ super
Yang-Mills theory (SYM) in four dimensions, has provided a most
promising framework. A lot of impressive results have been obtained
in this field based on the integrability discovered in the planar
limit of the ${\cal N}=4$ SYM. In particular, the thermodynamic
Bethe ansatz approach based on non-perturbative worldsheet
$S$-matrix has been formulated to provide the conformal dimensions
of SYM operators with arbitrary number of elementary fields for
generic value of 't Hooft coupling constant $\lambda$ (for a recent
review see  \cite{review}). In the strong coupling limit $\lambda\gg
1$, the AdS/CFT duality relates the conformal dimensions to the
energy and angular momenta of certain classical string
configurations.

The situation is similar for the $\gamma$-deformed (or
TsT-transformed) case, namely duality between string theory on
$AdS_5\times S^5_\gamma$ background, dual to $\mathcal{N} = 1$ SYM,
arising as marginal deformation of $\mathcal{N} = 4$ SYM
\cite{LM05,F05}.

There have been many interesting achievements on three-point
correlation functions in the AdS/CFT context. Three-point functions
for chiral primary operators have been computed first in the ${\rm
AdS}_5$ supergravity approximation \cite{sugra}. Recently, several
interesting developments have been made by considering general heavy
string states. An efficient method to compute two-point correlation
functions in the strong coupling limit is to evaluate string
partition function for a heavy string state propagating in the AdS
space between two boundary points based on a path integral method
\cite{Janik,BuchTsey}. This method has been extended to the
three-point functions of two heavy string states and a light
supergravity mode \cite{Zarembo,Costa,Roiban}. Relying on these
achievements, many interesting results concerning three-point
correlators of two heavy modes and one light mode have been obtained
\cite{Zarembo}-\cite{GG}.

The three-point functions of two heavy operators and a light
operator can be approximated by a supergravity vertex operator
evaluated at the heavy classical string configuration: \bea \nn
\langle V_{H}(x_1)V_{H}(x_2)V_{L}(x_3)\rangle=V_L(x_3)_{\rm
classical}. \eea For $\vert x_1\vert=\vert x_2\vert=1$, $x_3=0$, the
correlation function reduces to \bea \nn \langle
V_{H_1}(x_1)V_{H_2}(x_2)V_{L}(0)\rangle=\frac{C_{123}}{\vert
x_1-x_2\vert^{2\Delta_{H}}}. \eea Then, the normalized structure
constants \bea \nn \mathcal{C}_3=\frac{C_{123}}{C_{12}} \eea can be
found from \bea \label{nsc} \mathcal{C}_3=c_{\Delta}V_L(0)_{\rm
classical}, \eea were $c_{\Delta}$ is the normalized constant of the
corresponding light vertex operator.

Very recently, first results describing finite-size effects on the
three-point correlators appeared \cite{AB1105,Lee:2011,AB11062}. In
this note we are going to obtain more three-point correlation
functions for the cases when the heavy string states are {\it
finite-size} giant magnons with one or two angular momenta, and for
two different choices of the light state, corresponding to dilaton
operator and primary scalar operator. We first consider the case of
giant magnons in $AdS_5\times S^5$, and then we extend the obtained
results to giant magnons on $\gamma$-deformed $AdS_5\times
S^5_\gamma$ background.

\setcounter{equation}{0}
\section{$AdS_5\times S^5$ and $\mathcal{N} = 4$ SYM}
Let us first recall the relations between the coordinates, which we
are going to use further on. If we denote the string embedding
coordinates on $AdS$ and sphere parts of the $AdS_5\times S^5$
background with $Y$ and $X$ respectively, then \bea\nn
&&Y_1+iY_2=\sinh\rho\ \sin\eta\ e^{i\varphi_1},\h
Y_3+iY_4=\sinh\rho\ \cos\eta\ e^{i\varphi_2},\h Y_5+iY_0=\cosh\rho\
e^{it}, \eea are related to the Poincare coordinates by
 \bea \nn Y_m=\frac{x_m}{z},\h
Y_4=\frac{1}{2z}\left(x^mx_m+z^2-1\right), \h
Y_5=\frac{1}{2z}\left(x^mx_m+z^2+1\right), \eea where $x^m
x_m=-x_0^2+x_ix_i$, with $m=0,1,2,3$ and $i=1,2,3$.

Since we are going to compute the three-point correlators containing
two heavy operators corresponding to dyonic giant magnons, we
restrict ourselves to the $R_t\times S^3$ subspace of $AdS_5\times
S^5$. In that case, we can explore the reduction of the string
dynamics to the Neumann-Rosochatius (NR) integrable system by using
the ansatz \cite{KRT06} \bea\label{NRA}
&&t(\tau,\sigma)=\kappa\tau,\h \theta(\tau,\sigma)=\theta(\xi),\h
\phi_j(\tau,\sigma)=\omega_j\tau+f_j(\xi),\\ \nn
&&\xi=\alpha\sigma+\beta\tau,\h \kappa, \omega_j, \alpha,
\beta=constants,\h j=1,2.\eea Then the string Lagrangian in
conformal gauge, on the three-sphere, can be written as (prime is
used for $d/d\xi$) \bea\nn &&\mathcal{L}_{S^3}=(\alpha^2-\beta^2)
\left[\theta'^2+\sin^2\theta\left(f'_1-\frac{\beta\omega_1}{\alpha^2-\beta^2}\right)^2
+\cos^2\theta\left(f'_2-\frac{\beta\omega_2}{\alpha^2-\beta^2}\right)^2
\right.
\\ \label{rl} &&-\left.\frac{\alpha^2}{(\alpha^2-\beta^2)^2}
\left(\omega_1^2\sin^2\theta+\omega_2^2\cos^2\theta\right)
\right].\eea One can show that the first integrals of the equations
of motion for $f_j(\xi)$, $\theta(\xi)$, take the form \bea\nn
&&f'_1=\frac{\omega_1}{\alpha}\frac{v}{1-v^2} \left(\frac{
W}{1-\chi} -1\right),
\\ \label{tfi} &&f'_2=-\frac{\omega_1}{\alpha}\frac{uv}{1-v^2} ,
\\ \nn && \theta'
=\frac{\omega_1}{\alpha}\frac{\sqrt{1-u^{2}}}{1-v^2}
\sqrt{\frac{(\chi_{p}-\chi)(\chi-\chi_{m})}{1-\chi}},\eea where
\bea\nn &&\chi_p+\chi_m=\frac{2-(1+v^2)W-u^2}{1
-u^2},\\
\label{3eqs} &&\chi_p \chi_m=\frac{1-(1+v^2)W+(v W)^2}{1 -u^2}.\eea
The case of finite-size giant magnons, corresponds to \bea\nn
0<\chi_{m}<\chi< \chi_{p}<1.\eea In (\ref{tfi}) and (\ref{3eqs}) the
following notations have been used \bea\nn &&\chi=\cos^2\theta, \h
v=-\frac{\beta}{\alpha},\h u=\frac{\omega_2}{\omega_1},\h
W=\left(\frac{\kappa}{\omega_1}\right)^2 .\eea The replacement into
(\ref{rl}) gives (we set for simplicity $\alpha=\omega_1=1$ )
\bea\label{Lgm} &&\mathcal{L}_{S^3}^{gm}=
-\frac{1}{1-v^2}\left[2-(1+v^2)W
-2\left(1-u^2\right)\chi\right].\eea Let us also note that the first
integral for $\chi$ is given by \bea\label{fichi} \chi'=
\frac{2\sqrt{1-u^{2}}}{1-v^2}
\sqrt{\chi(\chi_{p}-\chi)(\chi-\chi_{m})}.\eea

\subsection{Dilaton operator}
For the dilaton vertex we have \cite{Roiban} \bea \label{dv}
V^d=\left(Y_4+Y_5\right)^{-\Delta_d}
\left(X_1+iX_2\right)^j\left[z^{-2}\left(\p_+x_{m}\p_-x^{m}+\p_+z\p_-z\right)
+\p_+X_{k}\p_-X_{k}\right], \eea where the scaling dimension
$\Delta_d=4+j$ to the leading order in the large $\sqrt{\lambda}$
expansion. The corresponding operator in the dual gauge theory
should be proportional to $Tr\left(F_{\mu\nu}^2 \
Z^j+\ldots\right)$, or for $j=0$, just to the SYM Lagrangian.

The normalized structure constant (\ref{nsc}) can be computed by
using (\ref{dv}), applied for the case of giant magnons, to be
\cite{Hernandez2,AB1105} ($i\tau=\tau_e$)\bea\label{c3d}
\mathcal{C}_3=c_{\Delta}^{d}\int_{-\infty}^{\infty}\frac{d\tau_e}{\cosh^{4+j}(\sqrt{W}\tau_e)}
\int_{-L}^{L}d\sigma\left(W +\mathcal{L}_{S^3}^{gm}\right).\eea
Here, the parameter $L$ is introduced to take into account the {\it
finite-size} of the giant magnons. The normalization constant of the
dilaton vertex operator $c_{\Delta}^{d}$ is given by
\cite{BCFM98,Zarembo} \bea\nn
c_{\Delta}^{d}=\frac{\sqrt{\lambda}}{128\pi N}
\frac{\sqrt{(j+1)(j+2)(j+3)}}{2^{j}}.\eea

The integration over $\tau_e$ in (\ref{c3d}) gives \bea\nn
\int_{-\infty}^{\infty}\frac{d\tau_e}{\cosh^{4+j}(\sqrt{W}\tau_e)}=
\sqrt{\frac{\pi}{W}}\
\frac{\Gamma\left(\frac{4+j}{2}\right)}{\Gamma\left(\frac{5+j}{2}\right)}.\eea
The integration over $\sigma$ can be replaced by integration over
$\chi$ according to \bea\label{int} \int_{-L}^{L}d\sigma=
2\int_{\chi_m}^{\chi_p}\frac{d\chi}{\chi'},\eea where $\chi'$ is
given in (\ref{fichi}). As a result (\ref{c3d}) becomes
\bea\label{c3df} &&\mathcal{C}_3=2\pi^{3/2}c_{\Delta}^{d}
\frac{\Gamma\left(\frac{4+j}{2}\right)}{\Gamma\left(\frac{5+j}{2}\right)}
\frac{\chi_p^{\frac{j-1}{2}}}{\sqrt{(1-u^2)W}}
\\ \nn &&\left[(1-u^2)\chi_p
\
{}_2F_1\left(\frac{1}{2},-\frac{1}{2}-\frac{j}{2};1;1-\frac{\chi_m}{\chi_p}\right)\right.
\\ \nn &&-\left.\left(1-W\right)
\
{}_2F_1\left(\frac{1}{2},\frac{1}{2}-\frac{j}{2};1;1-\frac{\chi_m}{\chi_p}\right)\right].\eea
A few comments are in order. The structure constant in (\ref{c3df})
corresponds to finite-size dyonic giant magnons, i.e. with two
angular momenta. The case of finite-size giant magnons with one
angular momentum nonzero can be obtained by setting $u=0$ ($\chi_p$,
$\chi_m$ also depend on $u$ according to (\ref{3eqs})). The infinite
size case \cite{Hernandez2} is reproduced for $W=1$, $\chi_m=0$. For
$j=0$, (\ref{c3df}) reduces to the result of \cite{AB1105}.

\subsection{Primary scalar operator}

The primary scalar vertex is \cite{BCFM98,Zarembo,Roiban} \bea
\label{prv} V^{pr}=\left(Y_4+Y_5\right)^{-\Delta_{pr}}
\left(X_1+iX_2\right)^j\left[z^{-2}\left(\p_+x_{m}\p_-x^{m}-\p_+z\p_-z\right)
-\p_+X_{k}\p_-X_{k}\right],\eea where now the scaling dimension is
$\Delta_{pr}=j$. The corresponding operator in the dual gauge theory
is $Tr\left( Z^j\right)$.

For giant magnons we have \cite{Hernandez2} \bea\nn
z^{-2}\left(\p_+x_{m}\p_-x^{m}-\p_+z\p_-z\right)=
\kappa^2\left(\frac{2}{\cosh^2(\kappa\tau_e)}-1\right).\eea Then the
light vertex operator becomes \bea\nn  V^{pr}=
\frac{\cos^j\theta}{\cosh^j(\kappa\tau_e)}
\left[\kappa^2\left(\frac{2}{\cosh^2(\kappa\tau_e)}-1\right)-\mathcal{L}_{S^3}^{gm}\right],\eea
where the infinite-size case was considered in \cite{Hernandez2},
while for the finite-size giant magnons $\mathcal{L}_{S^3}^{gm}$
should be taken from (\ref{Lgm}). As a consequence, the normalized
structure constant in the corresponding three-point function, for
the case under consideration, takes the form: \bea\label{c3pr}
\mathcal{C}_3^{pr}&=&c_{\Delta}^{pr}
\left[\int_{-\infty}^{\infty}d\tau_e
\frac{W}{\cosh^{j}(\sqrt{W}\tau_e)}
\left(\frac{2}{\cosh^2(\sqrt{W}\tau_e)}-1\right)
\int_{-L}^{L}d\sigma\chi^{\frac{j}{2}} \right.
\\ \nn &&-\left.\int_{-\infty}^{\infty} \frac{d\tau_e}{\cosh^{j}(\sqrt{W}\tau_e)}
\int_{-L}^{L}d\sigma\chi^{\frac{j}{2}}\mathcal{L}_{S^3}^{gm}\right].\eea

Performing the integrations in (\ref{c3pr}), by using (\ref{int}),
(\ref{fichi}), one finally finds \bea\label{c3prf}
&&\mathcal{C}_3^{pr}=\pi^{3/2}c_{\Delta}^{pr}
\frac{\Gamma\left(\frac{j}{2}\right)}{\Gamma\left(\frac{3+j}{2}\right)}
\frac{\chi_p^{\frac{j-1}{2}}}{\sqrt{(1-u^2)W}}
\\ \nn &&\left[\left(1-W+j(1-v^2W)\right)
\
{}_2F_1\left(\frac{1}{2},\frac{1}{2}-\frac{j}{2};1;1-\frac{\chi_m}{\chi_p}\right)\right.
\\ \nn &&-\left.\left(1+j\right)\left(1-u^2\right)\chi_p
\
{}_2F_1\left(\frac{1}{2},-\frac{1}{2}-\frac{j}{2};1;1-\frac{\chi_m}{\chi_p}\right)\right].\eea
As before, the case of finite-size giant magnons with one angular
momentum can be obtained by setting $u=0$ in (\ref{c3prf}).

It was observed in \cite{Hernandez2} that for the infinite-size
case, $\mathcal{C}_3^{pr}$ is nonzero only for giant magnons with
two angular momenta. For giant magnons with one angular momentum
$\mathcal{C}_3^{pr}$ vanishes. Let us see if we can reproduce this
result from (\ref{c3prf}). To this end, we fix $W=1$, $\chi_m=0$.
Then according to (\ref{3eqs}), $\chi_p$ becomes \bea\nn
&&\chi_p=\frac{1-v^2-u^2}{1-u^2}, \eea and one can show that
$\mathcal{C}_3^{pr}$ reduces to \bea\label{Cinf}
\mathcal{C}_{3\infty}^{pr}= 2\pi c_{\Delta}^{pr}
\frac{\Gamma\left(\frac{j}{2}\right)\Gamma\left(1+\frac{j}{2}\right)}
{\Gamma\left(\frac{3+j}{2}\right)\Gamma\left(\frac{1+j}{2}\right)}
\frac{u^2}{\sqrt{1-u^2-v^{2}}}
\left(\frac{1-v^2-u^2}{1-u^2}\right)^{j/2}.\eea Since we know that
the case of giant magnons with one angular momentum corresponds to
$u=0$, (\ref{Cinf}) confirms the observation made in
\cite{Hernandez2}.

It is also interesting to see how (\ref{Cinf}) looks written in
terms of the second angular momentum $J_2$ and the worldsheet
momentum $p$ of the string, where $p$ is identified with the magnon
momentum in the dual spin chain on the field theory side. For this
purpose, we have to know the relations between $(v,u)$ and
$(J_2,p)$. One way to find them is to look at the finite-size
expansions of $J_2$, $p$, and to take into account only the leading
terms there, because at the moment we are considering the
infinite-size case. According to \cite{PB10}, this gives
\bea\label{zms}
v=\frac{\sin(p)}{\sqrt{\mathcal{J}_2^2+4\sin^2(p/2)}},\h
u=\frac{\mathcal{J}_2}{\sqrt{\mathcal{J}_2^2+4\sin^2(p/2)}},\eea
 where \bea\nn \mathcal{J}_2=\frac{2\pi J_2}{\sqrt{\lambda}}
.\eea Replacing (\ref{zms}) into (\ref{Cinf}), one ends up with the
following expression for $\mathcal{C}_{3\infty}^{pr}$
\bea\label{Cinff} \mathcal{C}_{3\infty}^{pr}= \pi c_{\Delta}^{pr}
\frac{\Gamma\left(\frac{j}{2}\right)\Gamma\left(1+\frac{j}{2}\right)}
{\Gamma\left(\frac{3+j}{2}\right)\Gamma\left(\frac{1+j}{2}\right)}
\mathcal{J}_2^2\frac{\sin^{j-2}(p/2)}{\sqrt{\mathcal{J}_2^2+4\sin^2(p/2)}}.\eea

\setcounter{equation}{0}
\section{$AdS_5\times S^5_\gamma$ and $\mathcal{N} = 1$ SYM}

Investigations on AdS/CFT duality \cite{AdS/CFT} for the cases with
reduced or without supersymmetry is of obvious interest and
importance. An interesting example of such correspondence between
gauge and string theory models with reduced supersymmetry is
provided by an exactly marginal deformation of $\mathcal{N} = 4$
super Yang-Mills theory \cite{LS95} and string theory on a
$\beta$-deformed $AdS_5\times S^5$ background suggested by Lunin and
Maldacena in \cite{LM05}. When $\beta\equiv\gamma$ is real, the
deformed background can be obtained from $AdS_5\times S^5$ by the
so-called TsT transformation. It includes T-duality on one angle
variable, a shift of another isometry variable, then a second
T-duality on the first angle \cite{F05}.

An essential property of the TsT transformation is that it preserves
the classical integrability of string theory on $AdS_5\times S^5$
\cite{F05}. The $\gamma$-dependence enters only through the {\it
twisted} boundary conditions and the {\it level-matching} condition.
The last one is modified since a closed string in the deformed
background corresponds to an open string on $AdS_5\times S^5$ in
general.

The parameter $\tilde{\gamma}$, which appears in the string action,
is related to the deformation parameter $\gamma$ as \bea\nn
\tilde{\gamma}= \sqrt{\lambda}\ \gamma .\eea The effect of
introducing $\gamma$ on the field theory side of the duality is to
modify the super potential as follows \bea\nn W\propto
tr\left(e^{i\pi\gamma}\Phi_1\Phi_2\Phi_3-e^{-i\pi\gamma}\Phi_1\Phi_3\Phi_2\right).\eea
This leads to reduction of the supersymmetry of the SYM theory from
$\mathcal{N}=4$ to $\mathcal{N}=1$.

Since we are going to consider three-point correlation functions
with two vertices corresponding to dyonic giant magnon states, we
can restrict ourselves to the subspace $R_t\times S_{\gamma}^3$ of
$AdS_5\times S_{\gamma}^5$ background. Then one can show that by
using the ansatz (\ref{NRA}), the string Lagrangian in conformal
gauge, on the $\gamma$-deformed three-sphere $S_{\gamma}^3$, can be
written as \cite{AB11062} \bea\nn
&&\mathcal{L}_\gamma=(\alpha^2-\beta^2)
\left[\theta'^2+G\sin^2\theta\left(f'_1-\frac{\beta\omega_1}{\alpha^2-\beta^2}\right)^2
+G\cos^2\theta\left(f'_2-\frac{\beta\omega_2}{\alpha^2-\beta^2}\right)^2
\right.
\\ \label{r2} &&-\left.\frac{\alpha^2}{(\alpha^2-\beta^2)^2}G
\left(\omega_1^2\sin^2\theta+\omega_2^2\cos^2\theta\right)
+2\alpha\tilde{\gamma}G\sin^2 \theta \cos^2 \theta \frac{\omega_2
f'_1-\omega_1 f'_2}{\alpha^2-\beta^2}\right], \eea where \bea\nn
G=\frac{1}{1+\tilde{\gamma}^2\sin^2 \theta \cos^2 \theta}.\eea

In full analogy with the undeformed case, by using (\ref{r2}) and
the Virasoro constraints, one can find the following first integrals
\bea\nn &&f'_1=\frac{\Omega_1}{\alpha}\frac{1}{1-v^2} \left[\frac{v
W-u K}{1-\chi} -v(1-\tilde{\gamma}K)-\tilde{\gamma}u\chi\right],
\\ \label{tfid} &&f'_2=\frac{\Omega_1}{\alpha}\frac{1}{1-v^2} \left[\frac{K}{\chi}
-uv(1-\tilde{\gamma}K)-\tilde{\gamma}v^2W+\tilde{\gamma}(1-\chi)\right],
\\ \nn && \theta'
=\frac{\Omega_1}{\alpha}\frac{\sqrt{1-u^{2}}}{1-v^2}
\sqrt{\frac{(\chi_{p}-\chi)(\chi-\chi_{m})(\chi-\chi_{n})}{\chi(1-\chi)}},\eea
where \bea\nn &&\chi_p+\chi_m+\chi_n=\frac{2-(1+v^2)W-u^2}{1
-u^2},\\
\label{3eqsd} &&\chi_p \chi_m+\chi_p \chi_n+\chi_m
\chi_n=\frac{1-(1+v^2)W+(v W-u K)^2-K^2}{1 -u^2},\\ \nn && \chi_p
\chi_m \chi_n=- \frac{K^2}{1 -u^2}.\eea Now \bea\nn &&
u=\frac{\Omega_2}{\Omega_1},\h
W=\left(\frac{\kappa}{\Omega_1}\right)^2,\h
K=\frac{C_2}{\alpha\Omega_1},
\\ \nn &&\Omega_1=\omega_1\left(1+\tilde{\gamma}\frac{C_2}{\alpha\omega_1}\right), \h
\Omega_2=\omega_2\left(1-\tilde{\gamma}\frac{C_1}{\alpha\omega_2}\right)
,\eea while $\chi$ and $v$ are the same as before:
$\chi=\cos^2\theta$, $v=-\beta/\alpha$. The case of finite-size
giant magnons corresponds to \bea\nn 0<\chi_{m}<\chi< \chi_{p}<1,\h
\chi_{n}<0.\eea

Replacing (\ref{tfid}) and (\ref{3eqsd}) into (\ref{r2}), we find
the final form of the Lagrangian to be (we fix $\alpha=\Omega_1=1$)
\bea\label{Lg} &&\mathcal{L}_{\tilde{\gamma}}=
-\frac{1}{1-v^2}\left[2-(1+v^2)W-2\tilde{\gamma}K
-2\left(1-\tilde{\gamma}K-u\left(u-\tilde{\gamma}uK+\tilde{\gamma}vW\right)\right)\chi\right].\eea
After setting $\tilde{\gamma}=0$ in (\ref{Lg}) it coincides with
(\ref{Lgm}) as it should be. Now, the first integral for $\chi$
reads \bea\label{fichid} \chi'= \frac{2\sqrt{1-u^{2}}}{1-v^2}
\sqrt{(\chi_{p}-\chi)(\chi-\chi_{m})(\chi-\chi_n)}.\eea

\subsection{Dilaton operator}
To compute the normalized structure constant (\ref{nsc}) for the
case of two finite-size dyonic giant magnons living on the
$\gamma$-deformed three-sphere $S_{\gamma}^3$, we have to modify
(\ref{c3d}) by using (\ref{Lg}) \bea\label{c3dg} \mathcal{C}_3\to
\mathcal{C}_3^{\gamma}=
c_{\Delta}^{d}\int_{-\infty}^{\infty}\frac{d\tau_e}{\cosh^{4+j}(\sqrt{W}\tau_e)}
\int_{-L}^{L}d\sigma\left(W
+\mathcal{L}_{\tilde{\gamma}}\right).\eea Replacing the integration
over $\sigma$ according to (\ref{int}) and using (\ref{fichid}), one
finds \bea\label{c3dgf} &&\mathcal{C}_3^{\gamma}=
2\pi^{3/2}c_{\Delta}^{d}
\frac{\Gamma\left(\frac{4+j}{2}\right)}{\Gamma\left(\frac{5+j}{2}\right)}
\frac{\chi_p^{j/2}}{\sqrt{(1-u^2)W\left(\chi_p-\chi_n\right)}}
\\ \nn &&
\left\{\left[1-\tilde{\gamma}K-u\left(u+\tilde{\gamma}(vW-uK)\right)\right]\chi_p
F_1\left(1/2,1/2,-1-j/2;1;1-\epsilon,1-\frac{\chi_m}{\chi_p}\right)\right.
\\ \nn
&&-\left.\left(1-W-\tilde{\gamma}K\right)
F_1\left(1/2,1/2,-j/2;1;1-\epsilon,1-\frac{\chi_m}{\chi_p}\right)\right\},\eea
where \bea\label{de}
\epsilon=\frac{\chi_{m}-\chi_{n}}{\chi_{p}-\chi_{n}},\eea and
$F_1(a,b_1,b_2;c;z_1,z_2)$ is one of the hypergeometric functions of
two variables ($Appell F_1$). In writing (\ref{c3dgf}), we used the
following property of $F_1(a,b_1,b_2;c;z_1,z_2)$ \cite{PBM-III}
\bea\nn F_1(a,b_1,b_2;c;z_1,z_2)=
(1-z_1)^{-b_1}(1-z_2)^{-b_2}F_1\left(c-a,b_1,b_2;c;\frac{z_1}{z_1-1},\frac{z_2}{z_2-1}\right).\eea
Thus, the arguments $(1-\epsilon,1-\chi_m/\chi_p)\in (0,1)$. This
representation gives the possibility to use the defining series for
$F_1(a,b_1,b_2;c;z_1,z_2)$, \bea\nn
F_1(a,b_1,b_2;c;z_1,z_2)=\sum_{k_1=\ 0}^{\infty}\sum_{k_2=\
0}^{\infty}\frac{(a)_{k_1+k_2}(b_1)_{k_1}(b_2)_{k_2}}{(c)_{k_1+k_2}}
\frac{z_1^{k_1}z_2^{k_2}}{k_1 !\ k_2 !},\h \vert z_{1,2}\vert <1
,\eea in order to consider the limits $\epsilon\to 0$,
$\chi_m/\chi_p\to 0$ , or both. The small $\epsilon$ limit
corresponds to considering the leading finite-size effect, while
$\epsilon=0$, $\chi_m=0$, $\chi_n=0$, $K=0$, $W=1$, describes the
infinite-size case.

\subsection{Primary scalar operator}
Since the $\gamma$-deformation affects only the sphere, to compute
the three-point correlator of two finite-size giant magnons and the
primary scalar operator in the deformed case, we can use
(\ref{c3pr}), where $\mathcal{L}_{S^3}^{gm}$ must be replaced with
(\ref{Lg}), i.e. \bea\label{c3prg}
\mathcal{C}_{3\tilde{\gamma}}^{pr}&=&c_{\Delta}^{pr}
\left[\int_{-\infty}^{\infty}d\tau_e
\frac{W}{\cosh^{j}(\sqrt{W}\tau_e)}
\left(\frac{2}{\cosh^2(\sqrt{W}\tau_e)}-1\right)
\int_{-L}^{L}d\sigma\chi^{\frac{j}{2}} \right.
\\ \nn &&-\left.\int_{-\infty}^{\infty} \frac{d\tau_e}{\cosh^{j}(\sqrt{W}\tau_e)}
\int_{-L}^{L}d\sigma\chi^{\frac{j}{2}}\mathcal{L}_{\tilde{\gamma}}\right].\eea
Performing the integrations in (\ref{c3prg}) one obtains
\bea\label{c3prgf} &&\mathcal{C}_{3\tilde{\gamma}}^{pr}=
\pi^{3/2}c_{\Delta}^{pr}
\frac{\Gamma\left(\frac{j}{2}\right)}{\Gamma\left(\frac{1+j}{2}\right)}
\frac{(1-v^2)\chi_p^{j/2}}{\sqrt{(1-u^2)\left(\chi_p-\chi_n\right)}}
\\ \nn &&
\left\{\left[\sqrt{W}\frac{j-1}{j+1} +\frac{1}{\sqrt{W}(1-v^2)}
\left(2-(1+v^2)W-2\tilde{\gamma}K\right)\right]\right.
\\ \nn && \left.\times
F_1\left(1/2,1/2,-j/2;1;1-\epsilon,1-\frac{\chi_m}{\chi_p}\right)\right.
\\ \nn
&&- \left.\frac{2}{\sqrt{W}(1-v^2)}\left[1-\tilde{\gamma}K
-u\left(u-\tilde{\gamma}u
K+\tilde{\gamma}vW\right)\right]\chi_p\right.
\\ \nn && \left. \times
F_1\left(1/2,1/2,-1-j/2;1;1-\epsilon,1-\frac{\chi_m}{\chi_p}\right)\right\}.\eea

It can be shown that (\ref{c3prgf}) reduces to the undeformed case
(\ref{c3prf}) if we fix \bea\nn \tilde{\gamma}=K=\chi_n=0\h
\Rightarrow \epsilon=\frac{\chi_m}{\chi_p}.\eea This can be done by
using the following property of the hypergeometric function $F_1$
\bea\nn F_1(a,b_1,b_2;c;z,z)={}_2F_1\left(a,b_1+b_2;c;z\right).\eea

For the infinite-size case, (\ref{c3prgf}) gives \bea\label{cginf}
&&\mathcal{C}_{3\tilde{\gamma}\infty}^{pr}= \pi c_{\Delta}^{pr}
\frac{\Gamma\left(\frac{j}{2}\right)\Gamma\left(1+\frac{j}{2}\right)}
{\Gamma\left(\frac{3+j}{2}\right)\Gamma\left(\frac{1+j}{2}\right)}
\mathcal{J}_2\frac{\sin^{j-2}(p/2)}{\sqrt{\mathcal{J}_2^2+4\sin^2(p/2)}}
\left[\mathcal{J}_2+\tilde{\gamma}\sin^2(p/2)\sin(p)\right].\eea

Obviously, for $\tilde{\gamma}=0$, (\ref{cginf}) reduces to
(\ref{Cinff}) as it should be. Also, (\ref{cginf}) shows that the
three-point correlator is zero even in the $\gamma$-deformed case
for $\mathcal{J}_2=0$, when the giant magnons are of infinite size.

\setcounter{equation}{0}
\section{Concluding Remarks}
In this paper we extended the recently obtained results concerning
finite-size effects on the three-point correlation functions in the
framework of the semiclassical approximation, corresponding to the
case when the heavy string states are {\it finite-size} giant
magnons with one or two angular momenta
\cite{AB1105,Lee:2011,AB11062} in several ways.

First of all, we considered the more general case of dilaton vertex
with nonzero Kaluza-Klein momentum $j\ne 0$, for which only the
infinite-size case was studied \cite{Hernandez2}. The investigations
on finite-size effects in \cite{AB1105,Lee:2011} were restricted to
the particular case of $j= 0$. In that case, the coupling is just to
the Lagrangian \cite{Zarembo,Costa,Roiban}, i.e. it corresponds to a
marginal deformation of the SYM two-point functions by the
Lagrangian.

Next, we generalized the infinite-size result of \cite{Hernandez2}
about two heavy giant magnon states and primary scalar operator to
take into account the finite-size effect.

On the third place, we extended the infinite-size result of
\cite{AR1106} and the finite-size result of \cite{AB11062}, by
considering $j\ne 0$.

Finally, we derived the finite-size effect on the normalized
structure constant in the three-point correlation function of two
heavy giant magnon states and primary scalar operator on the
$\gamma$-deformed $AdS_5\times S^5_\gamma$ background. Let us point
out that even the infinite-size case was not considered by now.

Obviously, many interesting questions are waiting to be answered in
this field of research.

\section*{Acknowledgements}
This work was supported in part by DO 02-257 grant.


\begin{thebibliography}{99}

\bibitem{BPZ} A. A. Belavin, A. M. Polyakov and A. B . Zamolodchikov,
``Infinite conformal symmetry in two-dimensional quantum field
theory'', Nucl . Phys. {\bf B241} 333 (1984).

\bibitem{AdS/CFT} J. M. Maldacena, ``The large N limit of superconformal
field theories and supergravity'', Adv. Theor. Math. Phys. {\bf 2},
231 (1998) [{arXiv:hep-th/9711200}];\\S. S. Gubser, I. R. Klebanov
and A. M. Polyakov, ``Gauge theory correlators from non-critical
string theory'',
Phys. Lett. {\bf B428}, 105 (1998) [{arXiv:hep-th/9802109}];\\
E. Witten, ``Anti-de Sitter space and holography'', Adv. Theor.
Math. Phys. {\bf 2}, 253 (1998) [{arXiv:hep-th/9802150}].

\bibitem{review} N. Beisert et.al., ``Review of AdS/CFT Integrability: An Overview,
[arXiv:hep-th/10123982v3].

\bibitem{LM05} O. Lunin and J. Maldacena,
``Deforming field theories with $U(1)\times U(1)$ global symmetry
and their gravity duals'', JHEP {\bf 0505} 033 (2005),
[arXiv:hep-th/0502086].

\bibitem{F05} S. Frolov, ``Lax pair for strings in Lunin-Maldacena background'',
JHEP {\bf 0505} 069 (2005), [arXiv:hep-th/0503201].

\bibitem{sugra}
D. Z. Freedman, S. D. Mathur, A. Matusis and L. Rastelli,
``Correlation functions in the CFTd/AdSd+1 correspondence'', Nucl.
Phys. {\bf B546} 96 (1999) 96 [arXiv:hep-th/9804058]. G. Chalmers,
H. Nastase, K. Schalm and R. Siebelink, ``R-current correlators in N
= 4 super Yang-Mills theory from anti-de Sitter supergravity'',
Nucl. Phys. {\bf B540} 247 (1999) [arXiv:hep-th/9805105]. S. Lee, S.
Minwalla, M. Rangamani and N. Seiberg, ``Three-point functions of
chiral operators in D = 4, N = 4 SYM at large N'', Adv. Theor. Math.
Phys. {\bf 2} 697 (1998) [arXiv:hep-th/9806074]. G. Arutyunov and S.
Frolov, ``Some cubic couplings in type IIB supergravity on AdS5× S5
and three-point functions in SYM(4) at large N'', Phys. Rev. {\bf
D61} 064009 (2000) [arXiv:hep-th/9907085].
\bibitem{Janik} R. A. Janik, P. Surowka and A. Wereszczynski,
``On correlation functions of operators dual to classical spinning
string states'', JHEP {\bf 1005} 030 (2010)
[arXiv:hep-th/1002.4613].

\bibitem{BuchTsey} E. I. Buchbinder and A. A. Tseytlin,
``On semiclassical approximation for cor-relators of closed string
vertex operators in AdS/CFT'', JHEP {\bf 1008} 057 (2010)
[arXiv:hep-th/1005.4516].

\bibitem{Zarembo} K. Zarembo, ``Holographic three-point functions of semiclassical states'',
JHEP {\bf 1009} 030 (2010) [arXiv:hep-th/1008.1059].
\bibitem{Costa} M. S. Costa, R. Monteiro, J. E. Santos and D. Zoakos,
``On three-point correlation functions in the gauge/gravity
duality'', JHEP {\bf 1011} 141 (2010) [arXiv:hep-th/1008.1070].
\bibitem{Roiban} R. Roiban and A. A. Tseytlin,
``On semiclassical computation of 3-point functions of closed string vertex operators in AdS5 × S5'',
Phys. Rev. {\bf D82} 106011 (2010) [arXiv:hep-th/1008.4921].

\bibitem{Hernandez:2010}
  R.~Hern\'andez,
  ``Three-point correlation functions from semiclassical circular
  strings'',
  J. Phys. A {\bf 44}, 085403 (2011)
  [arXiv:1011.0408 [hep-th]].

\bibitem{Ryang:2010}
  S.~Ryang,
  ``Correlators of Vertex Operators for Circular Strings with Winding Numbers in $AdS_5\times
S^5$'',
  JHEP {\bf 1101}, 092 (2011)
  [arXiv:1011.3573 [hep-th]].

\bibitem{Georgiou:2010}
  G.~Georgiou,
  ``Two and three-point correlators of operators dual to folded string solutions at strong
  coupling'',
  JHEP {\bf 1102}, 046 (2011)
  [arXiv:1011.5181 [hep-th]].

\bibitem{Russo:2010}
  J.~G.~Russo and A.~A.~Tseytlin,
  ``Large spin expansion of semiclassical 3-point correlators in $AdS_5\times
S^5$'',
  JHEP {\bf 1102}, 029 (2011)
  [arXiv:1012.2760 [hep-th]].

\bibitem{Park:2010}
  C.~Park and B.~Lee,
  ``Correlation functions of magnon and spike'',
  [arXiv:1012.3293 [hep-th]].

\bibitem{BT:2010}
  E.~I.~Buchbinder and A.~A.~Tseytlin,
  ``Semiclassical four-point functions in $AdS_5\times
S^5$'',
  JHEP {\bf 1102}, 072 (2011)
  [arXiv:1012.3740 [hep-th]].

\bibitem{Bak:2011}
  D.~Bak, B.~Chen and J.~Wu,
  ``Holographic Correlation Functions for Open Strings and Branes'',
  JHEP {\bf 1106}, 014 (2011)
  [arXiv:1103.2024 [hep-th]].

\bibitem{Bissi:2011}
  A.~Bissi, C.~Kristjansen, D.~Young and K.~Zoubos,
  ``Holographic three-point functions of giant gravitons'',
  [arXiv:1103.4079 [hep-th]].

\bibitem{Arnaudov:2011a}
  D.~Arnaudov, R.~C.~Rashkov and T.~Vetsov,
  ``Three- and four-point correlators of operators dual to folded string solutions in
  $AdS_5\times S^5$'',
  [arXiv:1103.6145 [hep-th]].

\bibitem{Hernandez2} R. Hern\'andez, ``Three-point correlators for giant
magnons'', JHEP {\bf 1105} 123 (2011) [arXiv:hep-th/1104.1160].

\bibitem{Bai:2011}
  X.~Bai, B.~Lee and C.~Park,
  ``Correlation function of dyonic strings'',
  [arXiv:1104.1896 [hep-th]].

\bibitem{Alday:2011}
  L.~F.~Alday and A.~A.~Tseytlin,
  ``On strong-coupling correlation functions of circular Wilson loops and local
  operators'',
  [arXiv:1105.1537 [hep-th]].

\bibitem{AB1105} C. Ahn and P. Bozhilov, ``Three-point Correlation functions of Giant magnons
with finite size'' [arXiv:hep-th/1105.3084v1].

\bibitem{Lee:2011}
  B.~Lee and C.~Park,
  ``Finite size effect on the magnon's correlation functions'',
  [arXiv:1105.3279 [hep-th]].

\bibitem{Klose:2011}
  T.~Klose and T.~McLoughlin,
  ``A light-cone approach to three-point functions in $AdS_5\times
  S^5$'',
  [arXiv:1106.0495 [hep-th]].

\bibitem{AR1106} D. Arnaudov, R.C. Rashkov,
``Three-point correlators: examples from Lunin-Maldacena
background'' [arXiv:hep-th/11064298].

\bibitem{AB11062} C. Ahn and P. Bozhilov, ``Three-point Correlation function of Giant Magnons
in the Lunin-Maldacena background'' [arXiv:hep-th/1106.5656v1].

\bibitem{GG} G. Georgiou, ``$SL(2)$ sector: weak/strong coupling agreement of three-point correlators'',
[arXiv:hep-th/1107.1850v1].

\bibitem{KRT06} M. Kruczenski, J. Russo, A.A. Tseytlin,
``Spiky strings and giant magnons on S5'', JHEP {\bf 0610} 002
(2006), [arXiv:hep-th/0607044v3]


\bibitem{AAF05} L. F. Alday, G. Arutyunov, S. Frolov,
``Green-Schwarz strings in TsT-transformed backgrounds'', JHEP {\bf
0606} 018 (2006), [arXiv:hep-th/0512253].

\bibitem{HS08} Y. Hatsuda and R. Suzuki, ``Finite-size effects from giant magnons'',
Nucl. Phys. {\bf B800} 349 (2008) [arXiv:hep-th/0801.0747].

\bibitem{BF08} D. Bykov and S. Frolov, ``Giant magnons in TsT-transformed
$AdS_5 \times S^5$'', JHEP {\bf 0807} 071 (2008)
[arXiv:hep-th/0805.1070v2].

\bibitem{AB2010} Changrim Ahn and P. Bozhilov, ``Finite-Size Dyonic Giant Magnons in
TsT-transformed $AdS_5\times S^5$'', JHEP {\bf 1007} 048 (2010),
[arXiv:hep-th/1005.2508v1]

\bibitem{BCFM98} D. Berenstein, R. Corrado, W. Fischler, J.
Maldacena, ``The Operator Product Expansion for Wilson Loops and
Surfaces in the Large N Limit'', Phys. Rev. {\bf D59} 105023 (1999)
[arXiv:hep-th/9809188v1].

\bibitem{PB10} P. Bozhilov, ``Close to the Giant Magnons'',
[arXiv:hep-th/1010.5465v1]


\bibitem{LS95} R. G. Leigh and M. J. Strassler,
``Exactly marginal operators and duality in four-dimensional
$\mathcal{N}=1$ supersymmetric gauge theory'', Nucl. Phys. {\bf B
447} 95 (1995), [arXiv:hep-th/9503121].

\bibitem{PBM-III}A.P. Prudnikov, Yu.A. Brychkov and O.I. Marichev, Integrals and
series. v.3: More special functions, NY, Gordon and Breach (1990).


\end{thebibliography}
\end{document}